\documentclass{iau} 
\usepackage{graphicx}

\def\apj{ApJ}%
\def\mnras{MNRAS}%
\title[Radio transients investigation with VLBI] 
{Radio transients investigation with VLBI}

\author[Sokolovsky, Giroletti, Corbel, Anderson \& Stappers]   
{K.~V.~Sokolovsky$^{1,2,3}$,
M.~Giroletti$^{4}$,
S.~Corbel$^{5,6}$,
G.~E.~Anderson$^{7}$,
B.~W.~Stappers$^{8}$}

\affiliation{$^{1}$IAASARS, National Observatory of Athens, 15236 Penteli, Greece \\[\affilskip]
$^{2}$Sternberg Astronomical Inst. MSU, Universitetskii~pr. 13, 119992 Moscow, Russia \\[\affilskip]
$^{3}$Astro Space Center, LPI RAS, Profsoyuznaya Str. 84/32, 117997 Moscow, Russia \\[\affilskip]
$^{4}$INAF, Istituto di Radio Astronomia di Bologna, Via~P.~Gobetti~101, 40129~Bologna, Italy \\[\affilskip]
$^{5}$Laboratoire AIM (CEA/IRFU - CNRS/INSU - Universit\'e Paris Diderot), CEA DSM/IRFU/DAp, F-91191 Gif-sur-Yvette, France. \\[\affilskip]
$^{6}$Station de Radioastronomie de Nan\c{c}ay, Observatoire de Paris, PSL Research University, CNRS, Univ. Orl\'{e}ans, 18330 Nan\c{c}ay, France \\[\affilskip]
$^{7}$ICRAR, Curtin University, GPO Box U1987, Perth, WA 6845, Australia \\[\affilskip]
$^{8}$Jodrell Bank Centre for Astrophysics, School of Physics and Astronomy, The University of Manchester, Manchester M13 9PL, UK}

\pubyear{2018}
\volume{339}  
\jname{Southern Horizons in Time-Domain Astronomy}
\editors{E. Griffin, R. Seaman, M. Sullivan \& P. Woudt eds.}
\begin{document}

\maketitle

\begin{abstract}
The technique of Very Long Baseline Interferometry (VLBI) can provide accurate
localization and unique physical information about radio transients.
However, it is still underutilized due to the inherent difficulties of VLBI data
analysis and practical difficulties of organizing observations on short notice.
We present a brief overview of the currently available VLBI arrays and observing strategies used to  
study long- and short-duration radio transients.
\end{abstract}

\firstsection 

\section{Introduction}

The Very Long Baseline Interferometry (VLBI; \cite{1999ASPC..180..433W}) technique combines signals recorded at distant radio telescopes to achieve the highest angular resolution. 
A typical VLBI scale of 1\,mas by definition corresponds to the linear size of 1\,AU at the distance of 1\,kpc and 1\,pc at $z \sim 0.05$.
%
%
The longer the baseline (distance) between the
elements, the higher is the interferometer's angular resolution. Another way
to increase angular resolution is to observe at a shorter wavelength.
The measured interferometer response may be compared to a simple model in order to estimate the source size and flux density or, if measurements at many baselines are available, the source image may be reconstructed.
%
The following features of VLBI may provide insights into the nature of various astrophysical transients:
\begin{itemize}
\item Superb angular resolution helps to measure the source size.
\item Accurate localization of the radio emitting site is possible with VLBI.
\item Imaging reveals the radio emitting region geometry (jet/shell/shock front) and allows us to follow its changes (proper motion, expansion).
\item Full Stokes imaging may provide clues about the mechanism responsible for the transient's radio emission and, in the case of synchrotron transients, measure the magnetic field strength and structure.
\item VLBI can separate the (small) transient source from the unrelated background emission that will be ``resolved out'', no matter how bright the background is.
\end{itemize}
In Section\,\ref{sec:arrays} we provide an overview of VLBI arrays performing astronomical observations. Section\,\ref{sec:transientzoo} lists the various types of radio transients and highlights selected observational results.
In Section\,\ref{sec:strategy} we discuss observing strategies suitable for transient source studies with VLBI.
This is part two of the Workshop on radio transients. The first part of the workshop highlighting open questions in the transients science and how they may be addressed with non-VLBI techniques is presented by Anderson~et~al. (these proceedings).

\section{An overview of VLBI arrays}
\label{sec:arrays}

The majority of arrays listed in this section offer at least part of their observing time as ``open sky'' (any astronomer can apply) and accept target of opportunity requests. A number of VLBI-capable telescopes are not part of these arrays: they are dedicated to either space geodesy (\cite{VLBIGeodynamics}) or deep space communication. 


 {\it The Very Long Baseline Array} (VLBA; \cite{1994IAUS..158..117N}) is the
first instrument fully dedicated to VLBI. It includes ten 25\,m telescopes
spread across the continental United States, US Virgin Islands and Hawaii.
It operates full time at frequencies 0.3--96\,GHz and is frequency agile,
meaning that it may switch between the receivers in about a minute.
The VLBA may be combined with the GBT~100\,m, phased VLA 27x25\,m, Arecibo~305\,m, Effelsberg~100\,m and/or the LMT~50\,m to form {\it the High Sensitivity Array }.

 {\it The European VLBI Network} (EVN; \cite{2015arXiv150105079Z}) is a
collaboration of 10--15 diverse stations (including 60--100\,m class
telescopes). The number of participating stations depends on the observing
band (1--43\,GHz range) and station availability. 
Most EVN stations are not frequency agile. Observations are performed 
during three session per year. There is a limited number of pre-planned
out-of-session observations. 
The EVN routinely includes stations from the regional VLBI arrays of Korea, Italy, China and Russia. 
The EVN may be requested together with the US stations as {\it the Global array}.

 {\it e-EVN} is a subset of the EVN capable of real time correlation.
This feature was specifically introduced for transient observations (\cite{2016arXiv161200508P}).
There is one 24\,hr e-EVN observing session per month. Additional ToO
observations are possible. 

 {\it The Global mm-VLBI Array} (GMVA; \cite{2014arXiv1407.8112H}) includes 
Effelsberg~100\,m, GBT~100\,m, 
NOEMA interferometer 7x15\,m, 
VLBA and 
other mm-band telescopes in Europe.
The observations are performed at 86\,GHz during two
sessions per year. 

 {\it The Event Horizon Telescope} (EHT; e.g. \cite{2014ApJ...788..120L}) is
a heterogeneous VLBI array observing at 230\,GHz. The EHT has one observing session per
year. The first open call for proposals for VLBI observations with the EHT together with the Atacama Large Millimeter/submillimeter Array is issued in 2018.

 {\it RadioAstron} (\cite{2013ARep...57..153K}) combines ground stations with the 10\,m radio telescope aboard the dedicated satellite 
 in a high elliptical orbit (apogee -- 326000\,km) to form a Space--VLBI array. The observing frequencies are 0.3, 1.7, 4.8, 22\,GHz. Observations at 22\,GHz may reach a higher angular resolution (\cite{2016ApJ...817...96G}) than the EHT at 230\,GHz. 
The first observation of a transient source with RaioAstron was the search for radio emission from SN2014J in M82 (\cite{2014ATel.6197....1S}).


 {\it The Long Baseline Array} (LBA; \cite{2015PKAS...30..659E}) has its
core stations in Australia (the largest are the ATCA interferometer 6x22\,m, Parkes~64\,m, and Tidbinbilla~70\,m)
but also provides intercontinental baselines to Hartebeesthoek~26\,m in South
Africa. This is the only VLBI array 
operating in Southern hemisphere.
The observing frequency range is 1.4--22\,GHz, but not all telescopes are available
at all bands. The observations are conducted in 3--4 sessions per year. 
Test observations of GRB~080409 combining a few LBA stations with the telescopes in China and Japan in the e-VLBI mode were performed by \cite{2016RAA....16..164M}.

 {\it The Korean VLBI Network} (KVN; \cite{2014AJ....147...77L}) consists of three dedicated 21\,m stations
capable of observing simultaneously at 22-43-86-130\,GHz (\cite{2013PASP..125..539H,2015AJ....150..202R}).
Possibilities of installing similar receiving systems at VLBI stations outside Korea are 
investigated by \cite{2015JKAS...48..277J}.

  {\it The VLBI Exploration of Radio Astrometry} (VERA; 
  \cite{2003ASPC..306..367K,2012IAUS..287..386H}) 
  array includes four 20\,m telescopes in Japan. Its main focus is on parallax and proper motions measurements of Galactic maser sources. VERA observes at 6.7 (methanol), 22 (water) and 43\,GHz (SiO masers) using the unique dual-beam system that allows simultaneous observations of the target maser source and an extragalactic continuum source serving as the phase calibrator. 

  {\it KaVa} combines KVN and VERA observing at 22 and 43\,GHz (e.g. \cite{2017PASJ...69...71H}).

  {\it The Italian VLBI network} 
  (\cite{2016ivs..conf..132S}) 
  includes the Sardinia\,64m and the two 32\,m telescopes at Medicina and Noto. It is capable of observing in the 1--22\,GHz range. \cite{2013ATel.5264....1S} searched for radio emission from SN2013ej in M74, using Medicina and Noto as a two-element VLBI.

  {\it The Japanese VLBI Network} (JVN; \cite{2006evn..confE..71D}) combines VERA with other VLBI-capable telescopes including Usuda 64\,m deep space communication antenna. No call for observing proposals from outside the JVN collaboration.

  {\it The Russian VLBI Network ``Quasar''} (\cite{2008evn..confE..53F}) includes three 32\,m telescopes in Svetloe, Zelenchukskaya and Badary observing in the 1--22\,GHz range. The main focus of the network is on geodetic VLBI, but it also performs astronomical observations with EVN and RadioAstron. There is no open call for proposals, but proposals for astronomical observations submitted directly to the director may be considered.

  {\it The Chinese VLBI Network} (CVN; \cite{2015IAUGA..2255896Z}) includes Tianma 65\,m, Miyun 50\,m, Kunming 30\,m and the 25\,m telescopes in Seshan and Urumqi. The network is used for spacecraft navigation, geodesy and astronomy. No open call for proposals.

  Future facilities include {\it the East Asia VLBI Network} (\cite{2016ASPC..502...81W,2018NatAs...2..118A}) that will combine the national networks of China, Japan and Korea and the African VLBI Network (\cite{2011saip.conf..473G,2016arXiv160802187C}).

 {\it LOFAR} with its international stations is a VLBI array 
operating at frequencies $\sim 50$ (\cite{2016MNRAS.461.2676M}) and $\sim150$\,MHz
(\cite{2016A&A...593A..86V}). Its angular resolution is comparable to that of connected interferometers operating at GHz frequencies.

 {\it e-MERLIN} (\cite{2009evlb.confE..29S}) is a 7-station (including Lovell 76\,m) array observing at 1--22\,GHz providing baselines 
approaching those of regional VLBI arrays, while technically 
being a connected interferometer. 
It was recently used to study Galactic transients, among others, by \cite{2014Natur.514..339C,2017MNRAS.469.3976H}.

\section{Types of radio transients}
\label{sec:transientzoo}


Radio-transients can be divided in two broad classes (\cite{2011BASI...39..353B}): 
{\it fast} transients likely related to neutron stars (and flares on low-mass stars) appear on sub-second
timescales and {\it slow} transients related to various explosive astrophysical events
that evolve on a timescale of days to months.
The fast transients include:

The enigmatic {\it Fast Radio Bursts} (FRB; \cite{2016PASA...33...45P}). Recent EVN$+$Arecibo observations allowed \cite{2017ApJ...834L...8M} to establish spatial coincidence of the repeating FRB~121102 with a persistent extragalactic radio source providing new constraints on the physical interpretation of the (repeating) FRB phenomenon. VLBI was used to investigate the suspected host 
of FRB~150418 (\cite{2016A&A...593L..16G,2016MNRAS.463L..36B}).

{\it Rotating radio transients} (RRATs; \cite{2015ApJ...809...67K}).

{\it Giant pulses from pulsars} (\cite{2012ApJ...760...64M}, \cite{2016PASP..128h4502T}).
%
The connection between the above three classes of fast transients is suspected (\cite{2018arXiv180100640P}), but not yet established.

{\it Flare stars} produce outbursts of non-thermal radio emission (\cite{2008ApJ...674.1078O}).

The following types of events may produce slow radio-transients:

   {\it Supernovae} are the most studied class of radio-transients (\cite{2017ARep...61..299B}).
    Over 50 radio supernovae are known (\cite{2011ApJ...740...23L}). 
    VLBI observations provide shell expansion velocity measurements 
    independent of optical spectroscopy and reveal 
    the mass-loss history of the progenitor star (e.g. \cite{2018MNRAS.475.1756B}).

   {\it $\gamma$-ray bursts} produce afterglows that may be detected (e.g. \cite{2013ApJ...779..105M,2016arXiv161006928M,2017A&A...598A..23N}) and 
   resolved (\cite{2007ApJ...664..411P}) with VLBI.

   {\it Novae and symbiotic stars} may appear as radio sources
    observable with VLBI, e.g. \cite{2008ApJ...685L.137S,2012evn..confE..47G}.
    The source of radio emission may be the nova shell and possibly
    non-relativistic synchrotron-emitting jet (\cite{2008ApJ...688..559R}).
    VLBI imaging of the $\gamma$-ray emitting classical nova V959\,Mon by \cite{2014Natur.514..339C} suggested the synchrotron emission is produced at the interface between the fast polar outflow and the slow thermally-emitting outflow escaping the binary system in the orbital plane. Understanding the structure of the shocks in nova ejecta is important as the shocks are found to be responsible not only for $\gamma$-ray, X-ray and radio (\cite{2016MNRAS.460.2687W}) but also contribute significantly to optical emission of novae (\cite{2017NatAs...1..697L}).

  {\it Dwarf novae} may also be transient radio sources (\cite{2008Sci...320.1318K}). 
     The mechanism of their radio emission is unclear.

  {\it Tidal disruption events (TDE)} in galactic nuclei,
   such as Swift~J164449.3$+$573451, may be detected in radio. This is
   interpreted as an evidence of a relativistic jet forming from the matter
   lost by the disrupted star (\cite{2012ApJ...748...36B}). Surprisingly, \cite{2016MNRAS.462L..66Y} where able to place the 
   upper limit of 0.3\,c on the ejection speed in this source.

   {\it Active galactic nuclei (AGN)} are known sources of variable radio
     emission and may appear as transients rising above
     the threshold of previous radio observations. 

   {\it Microquasars} (\cite{2010LNP...794...85G}) can flare by several orders of magnitude within days. Some of them are sources of radio emission also in the quiet state. Recent VLBI results include observations of the expanding jets in XTE\,J1908$+$094 by \cite{2017MNRAS.468.2788R} and the giant flare of Cygnus\,X-3 by \cite{2017MNRAS.471.2703E}.

   {\it Maser sources} associated with star forming regions and
late-type stars may show flares by orders of magnitude
(\cite{1988SvAL...14..468M,2007IAUS..242..330R}).

   {\it Other events}. Sometimes, even a combination of radio and
    multi-wavelength observations is not sufficient to determine the nature of a transient (e.g., \cite{2005Natur.434...50H}). 
    Such cases demand detailed investigation and can potentially lead to the understanding of novel astrophysical phenomena.


\section{Observing strategies}
\label{sec:strategy}

While, in principle, wide-field VLBI imaging (\cite{2013ApJ...768...12M}) may be used to {\it search} for slow transients, the most popular observing strategy so far is the {\it follow-up} 
of transients discovered at other wavelengths. 
The two key points to consider when planning observations are the array sensitivity and the possibility of rapid response.
A sensitive array includes big dishes and is capable of performing phase-referencing observations. Phase referencing 
makes the integration time (and hence the sensitivity) limited by the experiment duration rather than the atmosphere coherence time. Dedicated full-time arrays like VLBA and KVN, as well as ad~hoc arrays including only two to three telescopes can respond within days to a trigger (if the corresponding observing proposal is already in place). VLBI observations often rely on the Earth rotation to probe more spatial frequencies as the array elements move and improve the resulting image. This technique cannot be used for rapidly evolving transients. A ``snapshot'' observation will result in a degraded image (compared to a ``full track'' image) or may be suitable only for modeling, not image reconstruction. The quality of a snapshot image may be improved by adding more elements to the array.
Another point to consider for Galactic transients is their expected angular size. 
An explosive transient may take hours to days to reach the angular size of a few mas and become ``too big'' to be observed with VLBI. Unless it has a structure on smaller angular scales, it may be completely ``resolved-out'' by the interferometer.
The choice of observing frequency is less important then other considerations when observing synchrotron transients as they tend to have nearly flat spectra.

With the exception of repeating events like FRB~121102 or the ones possessing a long-term ``afterglow'', triggered observations of fast transients are not possible. Instead, the fast transients have to be found in the same data used to investigate them. 
Raw VLBI data (before being averaged in time and frequency by the correlator) are suitable for a fast transient search (\cite{2018AJ....155...98L}). The V-FASTR project is running a commensal survey for fast transients at the VLBA (\cite{2011ApJ...735...97W,2016PASP..128h4503W}).
One interesting possibility is shadowing a large single-dish telescope 
with a VLBI array, extending the observing strategy of \cite{2017ApJ...834L...8M} to a blind survey.

{\it The Square Kilometre Array} (SKA) will detect transients in real time providing targets for a VLBI follow-up.
Including the phased SKA into an existing VLBI network will boost the network sensitivity by more than an order of magnitude. This will enable detailed VLBI studies of the classes of sources that are now just barely detectable. Studies of classical VLBI targets such as AGNs will also benefit from access to a larger sample of observable sources an its extension towards low-luminosity objects. An overview of VLBI prospects for the SKA is presented by \cite{2015aska.confE.143P}, while \cite{2015aska.confE..53C} highlights perspectives for Galactic synchrotron transient studies.



\end{document}